\documentclass[twoside]{article}
\usepackage{fleqn,espcrc2}

\usepackage{graphicx}
\def\nuc#1#2{${}^{#1}$#2}
\def\per{Pb(ClO$_4$)$_2$}

\newcommand{\AmS}{{\protect\the\textfont2
  A\kern-.1667em\lower.5ex\hbox{M}\kern-.125emS}}
\title{Neutrino Detection using Lead Perchlorate}

\author{P. J. Doe, S. R. Elliott, C. Paul, R. G. H.
Robertson\address{University
of Washington, Seattle, Washington 98195, USA}}

\begin{document}
\begin{abstract}
We discuss the possibility of using lead perchlorate as a neutrino detector.
The  primary neutrino interactions are given along with some relevant
properties of  the material.
\vspace{1pc}
\end{abstract}
%
\maketitle
\section{Introduction}

Due to its large cross section and relative cheapness,  a number of groups
have expressed interest in using Pb as a target
for neutrino interactions to study supernovae\cite{LAND,OMNIS} or
oscillations\cite{HAR98}.  As a result there have been several
cross section calculations done recently\cite{LAND,FUL99,KOL99}.
The interesting neutrino interactions on Pb consist of:\\

$
\begin{array}{lclr}
\nu_e + ^{208}Pb & \Rightarrow & ^{208}Bi^* + e^- 	& (CC)\\
                        &             & \Downarrow & \\
                        &             & ^{207}Bi + x\gamma + yn & \\
\\
\nu_x + ^{208}Pb & \Rightarrow & ^{208}Pb^* + \nu_{x}^{'} & (NC) \\
                        &             & \Downarrow & \\
                        &             &^{207}Pb + x\gamma + yn & \\
\end{array}
$

The number of neutrons emitted (0, 1, or 2) depends on the neutrino energy
and whether the interaction is via the charged current (CC) or neutral
current (NC). The nuclear physics of this system is described in
Ref. \cite{FUL99}.

A lead-based neutrino detector must have an appreciable density of
Pb atoms and the capability of detecting the electrons, gammas and neutrons
produced in the reaction. Lead Perchlorate (\per) has a very high solubility
in water (500 g \per\,/100 g H$_{2}$O \cite{CRC}) and the saturated solution
appears transparent to the eye. The cost for an 80\% solution is approximately
\$10,000/tonne in quantities of 100 tonne \cite{GFS}. This raises the
possibility that a cost effective, Pb based, liquid $\check{C}$erenkov
detector can be constructed. The presence of \nuc{35}{Cl} provides a nucleus
with a high cross-section for neutron  capture with the subsequent emission
of  capture $\gamma$ rays totaling 8.4 MeV.

Using Pb as a target would make a powerful supernova detector
\cite{LAND,FUL99}. The average energies of neutrinos emitted by a supernova
are expected to follow a heirarchy: $E_{\nu_e} < E_{\bar{\nu}_e}
< E_{\nu_{\mu,\tau}}$ The  observation of high energy $\nu_e$ would be an
indication of ${\mu,\tau}$ oscillations.

The large cross section and delayed coincidence $\nu_e$ signature of Pb
could provide a high statistics oscillation experiment at a beam stop
\cite{KAR98} where a short duration beam spill such as at ISIS
allows the temparal separation of any mono energetic $\nu_e$ which result
from $\nu_{\mu}$ oscillation. The hydrogen content of \per\, solution also
makes
the detector sensitive to $\bar{\nu}_{\mu} \rightarrow \bar{\nu}_{e}$
oscillations.
Finaly, measuring the cross section for neutrino interactions in Pb is also
of importance to supernova modelers\cite{FUL99} investigating the explosion
mechanism and transmutation of nuclei.
\section{Physical Properties}
Some relevant properties of \per \, are given in Table 1.
\begin{table}
\caption{Some properties of an 80\% \per\, solution.}
\begin{tabular}{ll}	\hline
$^{208}$Pb number density	& 1.7 x $10^{21}$cm$^{-3}$	\\
$^{208}$Pb$(\nu_e, e^-)$ at 30 MeV	& 34.22 x $10^{-40}$cm$^{-3}$	\\
$^{35}$Cl n capture cross-section	& 44.0b	\\
Density	&	2.7 gm cm$^{-3}$	\\
Refractive index	& 1.50\\ \hline
\end{tabular}
\end{table}
To build a large $\check{C}$erenkov detector viewed by photo-multiplier tubes
from the periphery, the attenuation of the light must be minimal.
Data on the refractive index, spectral transmission and attenuation length
of various \per\, solutions were obtained using an 80\% solution from a
commercial source \cite{GFS}. No attempt to filter suspended particulates or
purify the solution was made. The strength of the solution was reduced using
deionized water with an attenuation length of greater than 20 meters.
The spectral  trasmission, referenced with respect to a deionized water sample
is given in Figure 2. There  are no obvious absorption regions seen in the
\per\,
sample between 300 and 600 nm, the sensitive region of most PMT's.
\begin{figure}
\begin{center}
\includegraphics[width=7.5cm]{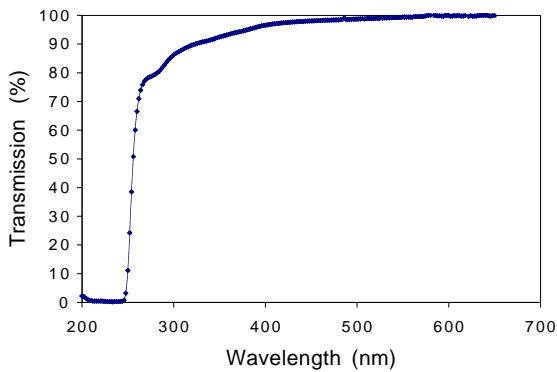}
\end{center}
\caption{Spectral transmission though a 1 cm cell of 80\% solution of \per\,
referenced to deionized water}
\end{figure}
Figure 3 shows the attenuation of light at 430 nm in an 80\% solution. These
data were obtained by passing a monochromatic, collimated beam of light
through a column of liquid and measuring the transmittance using a PMT.
The length of the column could be varied from 0 $\rightarrow$100 cm length
\begin{figure}
\begin{center}
\includegraphics[width=7.5cm]{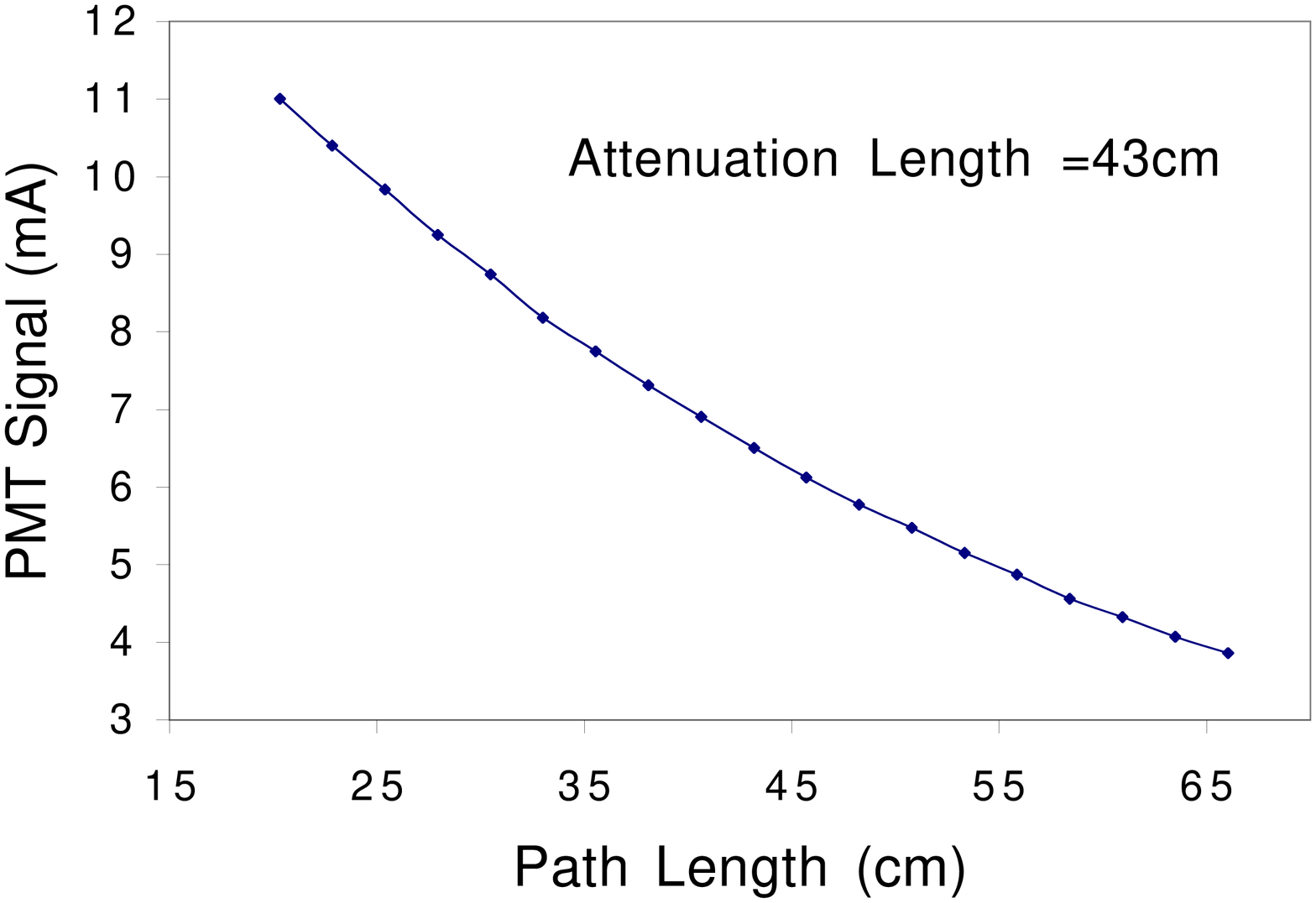}
\end{center}
\caption{Attenuation length of 430 nm light in 80\% \per. }
\end{figure}
\section{Discussion}
The lack of absorption lines in the transmission spectrum of \per\, is
encouraging. However,
the current attenuation length of 43 cm in an 80\% solution is too small to
realize a
conventional $\check{C}$erenkov neutrino detector. Further more, diluting
the solution with high
purity water resulted in significant reduction of the attenuation length
while the
transmission spectra were uneffected. This suggests that the loss of light
is due to scattering,
perhaps due to the formation of Pb salts or polymeric molecules such as
Pb$_4$(OH)$_4$,
possibly as a result of reaction with dissolved O$^2$ and CO$^2$.
The science of building massive, low background water $\check{C}$erenkov
detectors is well
understood. To demonstrate the feasibility of a \per\, $\check{C}$erenkov
detector, it
remains to investigate the chemistry pertinant to light transmission in the
solution.

\begin{thebibliography}{99}
\bibitem{LAND} C. K. Hargrove, {\it et al.}, Astroparticle Physics {\bf 5},
183 (1996).
\bibitem{OMNIS} D. B. Cline {\em et al.}, Phys. Rev. {\bf D50}, 720 (1994);
P. F. Smith, Astroparticle Physics {\bf 8}, 27 (1997).
\bibitem{HAR98} C. K. Hargrove, private communication.
\bibitem{FUL99} G. M. Fuller, W. C. Haxton, and G. C. McLaughlin,
Phys. Rev. D{\bf 59}, 085005 (1999).
\bibitem{KOL99} E. Kolbe, K. Langanke, and G. Mart$\acute{i}$nez-Pinedo,
Phys. Rev. C{\bf 60}, 052801 (1999).
\bibitem{CRC} Handbook of Chemistry and Physics, 65$^{th}$ edition, CRC
Press, Inc.
\bibitem{KAR98} B. Armbruster, {\it et al.}, Phys. Rev. {\bf C57}, 3414 (1998).
\bibitem{GFS} GFS Chemicals, P.O. Box 245, Powell, OH 43065.
\end{thebibliography}
\end{document}